\documentclass[aps,draft]{revtex4}
\usepackage[dvips]{graphicx}
\begin{document}

\title{Weibel Instabilities in a Completely Degenerate Electron Fermi Gas}
\author{ Levan N.Tsintsadze }
\affiliation{Department of Plasma Physics, E.Andronikashvili
Institute of Physics, Tbilisi, Georgia}

\date{\today}

\begin{abstract}

We study the Weibel instability in a degenerate Fermi plasma. We
disclose new type of quantum Weibel instabilities. In particular,
a novel oscillatory Weibel instability is found and its growth
rate is obtained. We reveal a transverse zero-sound in a quantum
degenerate electron gas, which has no counterpart in the classical
consideration.

\end{abstract}


\maketitle

In 1959, Weibel showed \cite{wei} that a self-excited transverse
electromagnetic (EM) waves can exist in an unmagnetized plasma,
when the electron distribution in the momentum space is
sufficiently anisotropic. The classical Weibel instability
\cite{fri}-\cite{laz} is a purely growing transverse EM mode
excited in an unmagnetized plasma by surplus the perpendicular
(relative to the propagation of EM waves) mean energy to the
parallel mean energy, i.e., $\varepsilon_\perp >
\varepsilon_\parallel$. In particular case it reads the
temperature anisotropy ($T_\perp > T_\parallel$).

The Weibel instability, which leads to the generation of
large-scale magnetic fields, has wide range of applicability to
astrophysical and space plasmas, as well as laboratory plasmas.
The relativistic theory was developed for it in
Refs.\cite{yo}-\cite{sch} in order to interpret the astrophysical
data on gamma-ray bursts, relativistic jets from active galactic
nuclei, and clusters of galaxies. The Wigner's type of quantum
kinetic equation for a plasma was derived by Klimontovich and
Silin \cite{kli}, and in the linear approximation obtained
dispersion relations for longitudinal plasma waves and transverse
EM waves in an isotropic plasma.

In this Brief Report, we consider the Weibel instability in fully
degenerate electron gas, supposing that a plasma consists of
degenerate electrons and nondegenerate ions (ions are assumed to
be immobile), such as those in neutron stars, supernovae, black
holes and white dwarfs. The purpose of the present work is to
investigate detailed properties of the Weibel instability in fully
degenerate electron plasma for a particular choice of anisotropic
distribution function.

As is well known in an electron gas at the temperature of absolute
zero (completely degenerate Fermi gas) the electrons occupy all
states with momentum from zero to $p=p_F=(3\pi^2)^{1/3}\hbar
n^{1/3}$, which is the radius of the Fermi sphere in momentum
space ($\hbar$ is the Planck constant divided by $2\pi$). Hence,
the possible electron trajectories in momentum space are closed.
In general the topology of Fermi surface can be more complex with
close electron trajectory \cite{lib,lifs}. As it was indicated in
Ref.\cite{lif} exists also the open Fermi surface, which means
that in some direction of momentum space the electron trajectory
is not closed. In our study we assume that the Fermi surface is
closed, but it has a cylindrical type of shape.

Let us consider one dimensional propagation of EM waves along the
z-axis. Relative to this propagation, we introduce the
perpendicular, $p_\perp$, and the parallel, $p_\parallel$,
momentums.

The distribution function adopted in the present calculation
assumes that some electrons occupy states with the perpendicular
momentum from zero to the limit $p_\perp=p_{F\perp}=\sqrt{
p_{Fx}^2+p_{Fy}^2}$, and some of the electrons occupy states with
parallel momentum in the range $p_\parallel=-p_{F\parallel}$ and
$p_\parallel=p_{F\parallel}$. So, the anisotropic distribution
function of completely degenerate electrons can be expressed as
\begin{eqnarray}
\label{dist} f_0=H(p_{F\perp}-p_\perp)\cdot
H(p_{F\parallel}^2-p_\parallel^2) \ ,
\end{eqnarray}
where $H(x)$ is the Heaviside step function, that is $H(x)=1$ for
$x>0$, and $H(x)=0$ for $x<0$.

The number of quantum states in an element of volume $dV$ with
values of electron momentum from $p_\perp$ to $p_\perp +dp_\perp$
and $p_\parallel$ to $p_\parallel +dp_\parallel$ equals $2\pi
p_\perp dp_\perp dp_\parallel dV/(2\pi\hbar)^3$. This quantum
number we must multiply by 2 due to the spin of electrons. Thus,
the N electrons, contained in the volume V, fill all the lowest
energy states with momentum ranging from zero to $p_{F\perp}$ and
$-p_{F\parallel}$ to $p_{F\parallel}$. Therefor, the density of
electrons can be presented as
\begin{eqnarray}
\label{dens} n=\frac{N}{V}=2\int_0^{p_{F\perp}}\frac{2\pi p_\perp
dp_\perp
}{(2\pi\hbar)^2}\int_{-p_{F\parallel}}^{p_{F\parallel}}\frac{
dp_\parallel}{2\pi\hbar} \ .
\end{eqnarray}
Based on the expression (\ref{dens}) we can introduce the number
of electrons that fill the energy states of perpendicular and
parallel directions
\begin{eqnarray}
\label{nel} n_\perp=\frac{p_{F\perp}^2}{2\pi^2\hbar^2}\ cm^{-2}
\hspace{1cm} and \hspace{1cm}
n_\parallel=\frac{p_{F\parallel}}{\pi\hbar}\ cm^{-1} \ .
\end{eqnarray}
With these expressions of $n_\perp$ and $n_\parallel$ at hand, we
write the kinetic energy as
\begin{eqnarray*}
\label{ken}
\varepsilon_{F\perp}=\frac{p_{F\perp}^2}{2m_0}=\frac{\pi^2\hbar^2}{m_0}
n_\perp \hspace{.5cm} and \hspace{.5cm}
\varepsilon_{F\parallel}=\frac{p_{F\parallel}^2}{2m_0}=
\frac{\pi^2\hbar^2}{2m_0}n_\parallel^2 \ ,
\end{eqnarray*}
where $m_0$ is the electron rest mass.

Correspondingly for the degeneracy temperatures we can write
\begin{eqnarray}
\label{dtem} T_{F\perp}=\frac{\varepsilon_{F\perp}}{K_B}
\hspace{1cm} and \hspace{1cm}
T_{F\parallel}=\frac{\varepsilon_{F\parallel}}{K_B} \ ,
\end{eqnarray}
where $K_B$ is the Boltzmann constant.

In order to investigate the Weibel instability at quantum scales,
we employ the expression of dielectric permeability,
$\varepsilon^{tr}(\omega,k)$, given in Ref.\cite{kuz}, which reads
\begin{eqnarray}
\label{per} \varepsilon^{tr}(\omega,k)=1-\frac{\omega_p^{2}}
{\omega^{2}}\Bigl(1-\frac{1}{\hbar m_0n_0}
\int\frac{p_\perp^2}{\omega -\vec{k}\cdot\vec{v}}\left[ f_0\Bigl(
\vec{p}+\frac{\hbar \vec{k}}{2}\Bigr)-f_0\Bigl(
\vec{p}-\frac{\hbar \vec{k}}{2}\Bigr) \right]\frac{d\vec{p}_\perp
dp_\parallel}{(2\pi\hbar)^3}\Bigr)\ ,
\end{eqnarray}
where $n_0$ is the total density of electrons and equals
$n_0=n_\perp\cdot n_\parallel$, $k$ is the wave-vector, $\omega$
is the wave frequency, $\omega_p$ is the electron plasma
frequency.

Using the dispersion relation $k^2c^2/\omega^2 = \varepsilon^{tr}$
($c$ is the speed of light in vacuum) for the transverse EM waves,
and taking into account the fact that we consider the propagation
of EM waves along the z-axis, we replace $\vec{p}+\hbar\vec{k}/2$
by the $\vec{p}$ in the first integral of Eq.(\ref{per}), and in
the second integral $\vec{p}-\hbar\vec{k}/2\rightarrow\vec{p}$,
and use the newly introduced distribution function (\ref{dist}) in
the expression (\ref{per}) to obtain
\begin{eqnarray}
\label{disp}\frac{k^2c^2}{\omega^2}=1-\frac{\omega_p^2}{\omega^2}
\left(1+\frac{
n_\perp}{kn_\parallel}\left[\ln\frac{\omega+\eta-kv_{F\parallel}}{
\omega+\eta+kv_{F\parallel}}-\ln\frac{\omega-\eta-kv_{F\parallel}}{
\omega-\eta+kv_{F\parallel}}\right]\right) \ ,
\end{eqnarray}
where $\eta = \hbar k^2/2m_0$. Note that the ratio
$\beta=n_\perp/kn_\parallel$ can in general be arbitrary.

We now examine the dispersion relation (\ref{disp}) for different
range of frequencies. First, for the weak quantum limit, when
$\mid\omega-kv_{F\parallel}\mid\gg\eta$, from Eq.(\ref{disp})
follows
\begin{eqnarray}
\label{dwq} \varepsilon^{tr}=1-\frac{\omega_p^2}{\omega^2}
\left(1+\frac{2n_\perp}{m_0n_\parallel}\frac{\hbar k\cdot
kv_{F\parallel}}{\omega^2-k^2v_{F\parallel}^2}\right) \ .
\end{eqnarray}
In the case $\omega^2\ll\omega_p^2$, expressing $v_{F\parallel}$
through $n_{F\parallel}$, we get the following dispersion equation
\begin{eqnarray}
\label{dise}\omega^2=k^2\frac{\pi^2\hbar^2}{4m_0^2}\left[
n_{F\parallel}^2-\frac{4}{\pi}\frac{\omega_p^2}{\omega_p^2+k^2c^2}
n_{F\perp}\right]\ .
\end{eqnarray}
We note here that if $\omega_p\sim kc$, then the condition for the
generation of low frequency EM field is
$n_{F\perp}>n_{F\parallel}^2$, which is similar to the one for the
classical Weibel instability, $T_\perp >T_\parallel$.

Next, we consider the range of frequencies $\mid\omega-\eta\mid\gg
kv_{F\parallel}$. In this case the dispersion relation
(\ref{disp}) reduces to
\begin{eqnarray}
\label{disr}
\omega^2=\eta\left(\eta-\frac{4\pi\omega_p^2}{\omega_p^2+k^2c^2}
\frac{\hbar n_\perp}{m_0}\right) \ .
\end{eqnarray}
Equation (\ref{disr}) predicts a purely growing quantum Weibel
instability for the wavelengths $\lambda\gg\sqrt{2\pi/n_\perp}$.
Here we assumed that $\omega_p^2\gg k^2c^2$.

In the frequency region $\mid\omega-\eta\mid <kv_{F\parallel} <
\omega+\eta$, use of the definition $\ln x=\ln\mid
x\mid-\imath\pi$ for $x<0$ in Eq.(\ref{disp}) yields
\begin{eqnarray}
\label{reg} \omega^2=\omega_p^2+k^2c^2+\imath\omega_p^2\pi\beta\ .
\end{eqnarray}
The solution of which we can write as
\begin{eqnarray}
\label{s11} \omega=\pm\sqrt{\omega_p^2+k^2c^2+\omega_p^2\pi\beta}
\left(\cos\frac{\varphi}{2}+\imath\sin\frac{\varphi}{2}\right)\ ,
\end{eqnarray}
where $\varphi={\rm
arctg}\frac{\omega_p^2\pi\beta}{\omega_p^2+k^2c^2}$.

If we now suppose that $\omega_p^2\pi\beta\ll\omega_p^2+k^2c^2$,
then $\cos\frac{\varphi}{2}\approx 1$ and
$\sin\frac{\varphi}{2}\sim\frac{\pi}{2}\beta
\frac{\omega_p^2}{\omega_p^2+k^2c^2}$, and from Eq.(\ref{s11})
follows that $Re\omega\gg Im\omega$, which indicates that there is
the possibility of an oscillatory Weibel instability in the
degenerate electron gas, that has no counterpart in classical
plasma. The growth rate of this novel oscillatory Weibel
instability is
\begin{eqnarray}
\label{gr}
Im\omega=\frac{\pi\omega_p^2\beta}{2\sqrt{\omega_p^2+k^2c^2}}\ .
\end{eqnarray}

In the opposite case, i.e.,
$\omega_p^2\pi\beta\gg\omega_p^2+k^2c^2$, one can immediately see
that $Re\omega\simeq Im\omega$ because $\varphi\sim\pi/2$,
$\cos\frac{\varphi}{2}\simeq\sin\frac{\varphi}{2}$, and we have
\begin{eqnarray}
\label{op} \omega\approx \omega_p\sqrt{\pi\beta}(1+\imath)\ .
\end{eqnarray}

Finally, we shall demonstrate the existence of transverse
zero-sound in a quantum degenerate electron gas. As is well known
in a Fermi liquid \cite{lifs}, as well as in a plasma the
longitudinal waves, known as the zero-sound exist in the range of
short wavelengths, i.e., $k^2r_D^2=k^2\frac{3v_F^2}{\omega_p^2}\gg
1$ ($r_D$ is the Debye length). Note that this wave is the
continuation of electron Langmuir waves. It should be emphasized
that the transverse zero-sound cannot arise in the classical
consideration. However, as we show here undamped transverse EM
zero-sound appears in quantum electron gas in the frequency region
$\omega\simeq kv_{F\parallel}-\eta$. To this end, we replace the
frequency by $\omega=kv_{F\parallel}-\eta+\gamma$ (where $\mid
kv_{F\parallel}-\eta\mid\gg\gamma$) in Eq.(\ref{disp}) to obtain
\begin{eqnarray}
\label{zs} \omega=kv_{F\parallel}-\eta\left(1+2\exp\Bigl\{-\frac{
k^2c^2}{\omega_p^2\beta}\Bigr\}\right) \ .
\end{eqnarray}
Note that for these waves to be true the inequalities
$k^2c^2\gg\omega_p^2\beta$ and $v_{F\parallel}\gg\hbar k/2m_0$
should hold. These inequalities can be satisfied in most practical
cases.

In summary, we have investigated the Weibel instability in fully
degenerate electron gas, assuming that a plasma consists of
degenerate electrons and nondegenerate ions, which are immobile. A
purely growing quantum Weibel instabilities, as well as
oscillatory Weibel instability are found. We have discovered the
transverse zero-sound in a quantum degenerate electron gas. We
specifically note here that this wave does not exist in classical
plasmas. The new quantum Weibel instabilities may be responsible
for the generation of non-stationary magnetic fields in compact
astrophysical objects as well as in the forthcoming intense
laser-solid density plasma experiments.

\end{document}